\documentclass[pra,groupedaddress,twocolumn,eqsecnum]{revtex4}

\usepackage{bm}
\usepackage{epsfig}

\begin{document}

\title{The Parity Non-Conserving $^3P_0 - ^1P_1$ E1 Transition
       Amplitude of the Atomic Yb }

\author{Angom Dilip Singh}
\email{angom@prl.ernet.in}
\affiliation{Physical Research Laboratory, Navarangpura, Ahmedabad-9, \\
             INDIA}
\author{Bhanu Pratap Das}
\email{das@iiap.ernet.in}
\affiliation{Non-Accelerator Particle Physics Group, Indian  Institute of 
             Astrophysics, \\
             Koramangala, Bangalore - 34, \\
             INDIA}

\begin{abstract}
   The atomic parity non-conservation(PNC) experiments has reached accuracies 
   which have important implications for physics beyond the standard model. An 
   optical rotation experiment to measure the atomic PNC of the 
   $^3P_1(6s6p) - ^1P_1(6s6p)$ in Yb was proposed 
   recently\cite{kimball-052113-2001}. Our screened electron-electron 
   coulomb potential multi-configuration Dirac-Fock calculation of th PNC 
   induced $E1$ transition amplitude of this transition
   $E1_{\rm PNC} = -96.0 \times 10^{-11}iea_0(-Q_W/N)$ of $^{171}{\rm Yb}$ is 
   more than two orders of magnitude larger than 
   $E1_{\rm PNC}(6s-7s) = -0.8991\times 10^{-11}iea_0(-Q_W/N)$ of 
   $^{133}{\rm Cs}$. 
\end{abstract}

\pacs{32.80.Ys, 31.30.Gs, 31.15.Ar}

\maketitle

%%%%%%%%%%%%%%%%%%%%%%%%%%%%%%%%%%%%%%%%%%%%%%%%%%%%%%%%%%%%%%%%%%%%%%%%%%%%%%%%
%%%%                              SECTION:1                                 %%%%
%%%%                            Introduction                                %%%%
%%%%%%%%%%%%%%%%%%%%%%%%%%%%%%%%%%%%%%%%%%%%%%%%%%%%%%%%%%%%%%%%%%%%%%%%%%%%%%%%

\section{Introduction}

   The nuclear weak charge $Q_W$ of an atom is a parameter of the 
atomic PNC arising from the nucleon-electron vector 
axial vector interaction. It is possible to extract $Q_W$ by measuring the 
PNC induced electric dipole transition amplitude $E1_{\rm PNC}$ and combining
it with atomic theory calculations. The $E1_{\rm PNC}$ of the $^{133}{\rm Cs}$ 
$6s-7s$ transition has been measured to very high accuracy\cite{wood-1759-97}. 
The $Q_W$ of $^{133}{\rm Cs}$ obtained from the experimental data after 
combining with the atomic theory 
calculations\cite{blundell-1411-90,dzuba-147-1989} was further refined  to
$72.06(28)_{\rm expt}(34)_{\rm theor}$ after the measurement of the $6s-7s$ 
transition polarizability\cite{bennet-2484-99}. This has a $2.5\sigma$
deviation from the predictions of the standard model
$Q_W=73.20(13)$\cite{marciano-2963-90}. The uncertainty of the experiment
results and atomic theory calculations are estimated at .35\% and 1\% 
respectively. This indicate the possibility of improving the results further
by reducing the atomic theory uncertainty. 

   The recent atomic theory calculations which treat the Breit interaction 
more rigorously have introduced corrections of .9\%\cite{derevianko-1618-2000},
.6\%\cite{dzuba-044103-2001} and .4\%\cite{kozlov-l607-2001} to $Q_W$ of 
$^{133}{\rm Cs}$. The variation of the results is due to the difference of the 
calculation methods and many-body effects included. The uncertainties of 
these calculations are estimated at 1\%. Considering the unique implications of 
the parameters obtained from the atomic PNC phenomena to the physics beyond the
standard model, it is desirable to reduce the atomic theory uncertainties and  
also confirm the $^{133}{\rm Cs}$ results using other atoms. 

 An important criteria of choosing a candidate atom is the 
$Z^3\alpha$ scaling of the PNC interaction, which indicates the advatage
of choosing high $Z$ atom. The $^1S_0 - ^3D_1$ transition of atomic 
Yb$(Z=70)$, which  has been studied 
theoretically\cite{das-1635-97,porsev-2781-99,
porsev-459-95,demille-4165-95} and experimentally\cite{bowers-3103-96,
bowers-3513-99} in detail is a promising candidate of the ongoing and future 
atomic PNC experiments. This transition is a suitable choice due to the
nearly degenerate $^3D_1(5d6s)(24489{\rm cm}^{-1})$ 
and $^1P_1(6s6p)(25068{\rm cm}^{-1})$, which can enhance the PNC mixing.  The  
PNC mixing enhancement between this pair of levels can also be exploited in 
the $^3P_0 - ^1P_1$ transition. 

  The $^3P_0 - ^1P_1$ transition was proposed for atomic
PNC experiment in a recent work\cite{kimball-052113-2001} and estimates
$E1_{\rm PNC}\approx 7\times 10^{-10}ea_0$ which is almost one order of 
magnitude larger than that of $^{133}{\rm Cs}$ 
$E1_{\rm PNC}\approx 8\times 10^{-11}ea_0$. The use of the metastable state
$^3P_0$ as the initial state offers the possibility of separating only 
the atomic density dependent plane of polarization rotation angle from the 
wave-length dependent systematic errors. This is expected to better the 
results obtained from the Stark-interference experiments, where the use
of the high intensity laser fields limits the statistical precision due to
the light shift.

  In this paper we have investigated the $E1_{\rm PNC}$ of the 
$^3P_{0,1} - ^1P_1$ transition using multi-configuration Dirac-Fock method. 
The atomic theory calculations are necessary to extract the nuclear weak 
charge $Q_W$ from the experimental value of $E1_{\rm PNC}$. To check the 
accuracy of the atomic calculations we also study the transition properties 
and hyper fine structure constants. All the calculations are in atomic units 
( $\hbar = e = m_e = 1$).

%%%%%%%%%%%%%%%%%%%%%%%%%%%%%%%%%%%%%%%%%%%%%%%%%%%%%%%%%%%%%%%%%%%%%%%%%%%%%%%%
%%%%                              SECTION:2                                 %%%%
%%%%                        Method of Calculation                           %%%%
%%%%%%%%%%%%%%%%%%%%%%%%%%%%%%%%%%%%%%%%%%%%%%%%%%%%%%%%%%%%%%%%%%%%%%%%%%%%%%%%

\section{The Method of Calculation}

  The nuclear spin-independent atomic PNC arises from the axial-vector vector 
electron-nucleon interaction component of the neutral weak current 
interaction between the electrons and nucleons of an atom. The interaction
is mediated by $Z_0$ bosons, which is a prediction of the electron-weak
unification. The effective form of the interaction Hamiltonian can be 
obtained by treating the nuclear part non-relativistically
\begin{equation}
   H_{\rm PNC}^{\rm NSI} = \frac{G_F}{\sqrt{8}}Q_W\gamma_5\rho_{\rm nuc}(r) ,
   \label{hpnc}
\end{equation}
where $G_F$ is the Fermi coupling constant, $\gamma_5$ is the Dirac matrix and
$\rho_{\rm nuc}$ is the nuclear density. The  PNC induced $E1$ transition 
amplitude between the initial and final states $|\Psi_i\rangle$ and 
$|\Psi_f\rangle$ is 
%\begin{widetext}
\begin{eqnarray}
   E1_{\rm PNC} & = & \sum_I\left [ \frac{\langle\Psi_f|\bm{D}|
                      \Psi_I\rangle\langle\Psi_I |H_{\rm PNC}^{\rm NSI}|
                      \Psi_i\rangle}{E_i - E_I} \right .\nonumber \\
                &   & \left . +\frac{\langle\Psi_f| H_{\rm PNC}^{\rm NSI}
                      |\Psi_I\rangle\langle\Psi_I|\bm{D}| 
                      \Psi_i\rangle}{E_f - E_I} \right ],
   \label{e1pnc}
\end{eqnarray}
%\end{widetext}
where $|\Psi_I\rangle$ are the intermediate states which are opposite in 
parity to $|\Psi_i\rangle$ and $|\Psi_f\rangle$, and $E_i$, $E_f$  and
$E_I$ are the energies of the states. 

  The numerator of (\ref{e1pnc}) has $H_{\rm PNC}^{\rm NSI}$, effective only 
within the nuclear region due to the nuclear density $\rho_{\rm nuc}(r)$ and 
the dipole $\bm{D}=-\bm{r}$ in length gauge, which has significant 
contribution from the large radial range. And the energies in the denominator 
has large contribution from the mid radial range where the electron density
is high. The calculation of the $E1_{\rm PNC}^{\rm NSI}$ require atomic state 
functions which are accurate over all radial ranges. $E1_{\rm PNC}$ can be 
calculated once the atomic wave-functions are known.

%%%%%%%%%%%%%%%%%%%%%%%%%%%%%%%%%%%%%%%%%%%%%%%%%%%%%%%%%%%%%%%%%%%%%%%%%%%%%%%%
%%%%                           SUBSECTION:2.1                               %%%%
%%%%                      Wave-function calculation                         %%%%
%%%%%%%%%%%%%%%%%%%%%%%%%%%%%%%%%%%%%%%%%%%%%%%%%%%%%%%%%%%%%%%%%%%%%%%%%%%%%%%%

\subsection{Wave-function Calculation}

  A method suitable for rare earth atoms which has large configuration mixing
is the multi-configuration Dirac-Fock (MCDF), which is the relativistic 
adaptation of the multi-configuration Hartree-Fock. The MCDF approximates an
atomic state function $|\Gamma PJM\rangle$ as a linear combination of 
configuration state functions (CSF) $|\gamma PJM\rangle$, which are again a 
linear combination of Slater determinants. Where $P$, $J$ and $M$ are the 
parity, total angular momentum and magnetic quantum numbers respectively, and 
$\Gamma$ and $\gamma$ are the additional quantum numbers required to defined 
the  ASF and CSFs uniquely. Then
\begin{equation}
   |\Gamma_i PJM\rangle = \sum_{r} c_{r\Gamma_i}|\gamma_r PJM\rangle ,
   \label{asf}
\end{equation}
where $c_{r\Gamma_i}$ are the coefficients of the CSFs. The energy functional
is defined using a set of ASFs which mix strongly and the Dirac-Coulomb
Hamiltonian
\begin{equation}
   H_{\rm DC} = \sum_i^{N} \left [ c \bm{\alpha}_i\cdot
                \bm{p}_i + \beta_i c^2 - \frac{Z}{r_i} \right ] +
                \sum_{i<j}^{N,N}\frac{1}{r_{ij}},
   \label{dcham}
\end{equation}
where $\bm{\alpha}_i$ and $\beta_i$ are the Dirac matrices, $\bm{p}_i$ is the 
momentum and $N$ is the number of electrons. The orbitals are then generated 
variationally. The orbitals from the negative continuum are also the solutions 
of such a method, however only the bound states can be chosen by imposing the 
boundary condition that the orbitals $\psi(\bm{r}) \rightarrow 0$ as 
$\bm{r}\rightarrow \infty$. Choosing the ASFs contributing to the 
valence-valence correlation effects, the orbitals captures important 
correlation effects. 

 Though MCDF method can represent the valence-valence correlation, it is not a
suitable method to calculate core-valence and core-core correlation effects.
A large number of CSFs is required to represent these correlation effects, 
which is computationally difficult as self consistent field method like MCDF 
is not suitable for a calculation involving large CSF spaces. However, these 
correlation effects can be calculated using configuration interaction(CI) 
using a set of virtual orbitals generated from the MCDF potential.

   A large set of CSFs is required in the CI calculation to represent all 
classes of correlation effects.  A more efficient method is to model it using 
the most important CSFs and modifying the form of the electron-electron Coulomb 
interaction potential as
\begin{equation}
   \frac{1}{r_{12}} = \sum_{K}\alpha_K\frac{r_<^K}{r_>^{K+1}}
                      \bm{C}_K(1) \cdot \bm{C}_K(2),
   \label{screen}
\end{equation}
where $\alpha_K$ are the constants that modifies the $K^{\rm th}$ multipole of 
the inter electron Coulomb interaction potential. The suitable values of these
constants can be obtained by matching properties like excitation energies 
calculated using this potential with the experimental data. The earlier 
calculations have shown that, this approach can reproduce the experimental
data to very good agreement\cite{das-1635-97,angom-4905-99}. The values of 
$\alpha$ are chosen such that $ 0. < \alpha_K < 1.$ and for consistency the
screening parameters are used in the MCDF calculations to generate the 
orbitals.

%%%%%%%%%%%%%%%%%%%%%%%%%%%%%%%%%%%%%%%%%%%%%%%%%%%%%%%%%%%%%%%%%%%%%%%%%%%%%%%%
%%%%                           SUBSECTION:2.2                               %%%%
%%%%                       Properties Calculation                           %%%%
%%%%%%%%%%%%%%%%%%%%%%%%%%%%%%%%%%%%%%%%%%%%%%%%%%%%%%%%%%%%%%%%%%%%%%%%%%%%%%%%

\subsection{Properties Calculation}

 The most important transition properties like oscillator strength, life time,
polarizability, etc arise from the electric dipole transition.  And all of 
these depend on the reduce matrix elements of the electric dipole operator. 
Consider the initial and final states of the atom $|\Psi_i\rangle$ and 
$|\Psi_f\rangle$, then from Wigner-Eckert theorem
\begin{equation}
   \bm{D}_{if} = \sum_q\left ( \begin{array}{ccc}
                        J_f &   1  & J_i  \\
                       -m_f &   q  & m_i  \\
                         \end{array}  \right ) (-1)^{(J_f - m_f)}
                         \langle \Psi_f||D||\Psi_i\rangle ,
   \label{wigner}
\end{equation}
where $J_f$ and $J_i$ are the total angular momenta of the final and initial
states, $m_f$ and $m_i$ are the magnetic quantum number of the final and 
initial states,  and $q$ is the component of the dipole operator. 
$\langle \Psi_f||D||\Psi_i\rangle = D_{fi}$ is the reduced matrix element and 
it is independent of geometry. The reduced matrix elements can be expressed 
in terms of the ASFs as
\begin{eqnarray}
   D_{fi} &=& \langle\Gamma_fP_fJ_f||D||\Gamma_iP_iJ_i\rangle \\
          &=& \sum_{rs}c_{r\Gamma_f}c_{s\Gamma_i}
              \langle\gamma_rP_fJ_f||D||\gamma_sP_iJ_i\rangle .
   \label{asfred}
\end{eqnarray}
The oscillator strength of the transition $|\Psi_i\rangle - |\Psi_f\rangle$ is 
\begin{equation}
    f_{fi} = \frac{2\Delta E}{3(2J_f+1)}|D_{fi}|^2 ,
    \label{osc}
\end{equation}
where $\Delta E = E_f - E_i$. The independent particle approximation of the 
inter-electron electromagnetic interaction implies that it is not gauge 
invariant. However, the gauge invariance is restored if the inter-electron 
electromagnetic interaction is represented completely by including the 
correlation effects correctly. The agreement of the dipole matrix elements
or transition properties calculated in different gauges indicate the 
completeness of the correlation effects included, which can also be 
interpreted as the accuracy of the wave-functions. 

 The hyperfine constants are the measure of the strength of the 
electron-nucleus parity allowed electromagnetic multipole interactions. The
general form of the interaction Hamiltonian is
\begin{equation}
   H_{\rm hfs} = \sum_{k} \bm{T}^{(k)}\cdot \bm {M}^{(k)},
   \label{hfsin}
\end{equation}
where $\bm{T}^{(k)}$ and $\bm {M}^{(k)}$ are spherical tensor operators of 
rank $k$ in the electron and nuclear space respectively. These represent the
electromagnetic multipoles. The $k = 1\,\,\text{and}\,\,2$ corresponding
to the magnetic dipole and electric quadrupole respectively are the most
important. The atomic states are then the eigenstates of the total
angular momentum
\begin{equation}
   \bm{F} = \bm{I} + \bm{J},
   \label{totang}
\end{equation}
where $\bm{I}$ is the nuclear spin. The shift in energy due to the magnetic
dipole and electric quadrupole hyperfine interactions of the atomic state
$|\gamma_I\gamma_JIJFM_f\rangle$ are
\begin{eqnarray}
   W_{M1}(J,J) &=& \frac{1}{2}A_JC        \nonumber\\
   W_{E2}(J,J) &=& B_J\frac{\frac{3}{4}C(C+1) - I(I+1)J(J+1)}
                   {2I(2I - 1)J(2J - 1)}  \nonumber
\end{eqnarray}
where $C = F(F+1) - J(J+1) - I I-1)$, and $A_J$ and $B_J$ are the magnetic
dipole and electric quadrupole hyperfine constants respectively. These are
defined as
\begin{eqnarray}
   A_J &=& \frac{\mu_I}{I}\frac{1}{[J(J+1)(2J+1)]^{\frac{1}{2}}}
           \langle\gamma_JJ||\bm{T}^{(1)}||\gamma_JJ\rangle    \nonumber \\
   B_J &=& 2Q\left [ \frac{2J(2J-1)}{(2J+1)(2J+2)(2J+3)}\right ]^{\frac{1}{2}}
           \langle\gamma_JJ||\bm{T}^{(2)}||\gamma_JJ\rangle    \nonumber
\end{eqnarray}
where $\mu_I$ and $Q$ are the nuclear magnetic dipole and electric quadrupole
moments.

%%%%%%%%%%%%%%%%%%%%%%%%%%%%%%%%%%%%%%%%%%%%%%%%%%%%%%%%%%%%%%%%%%%%%%%%%%%%%%%%
%%%%                            SECTION:3.1                                 %%%%
%%%%                Results: Screened Coulomb Potential                     %%%%
%%%%%%%%%%%%%%%%%%%%%%%%%%%%%%%%%%%%%%%%%%%%%%%%%%%%%%%%%%%%%%%%%%%%%%%%%%%%%%%%

\section{Results}

\subsection{Screened Coulomb Potential}

\begin{figure}[h]
   \epsfig{height=3.5in,width=3.5in,figure=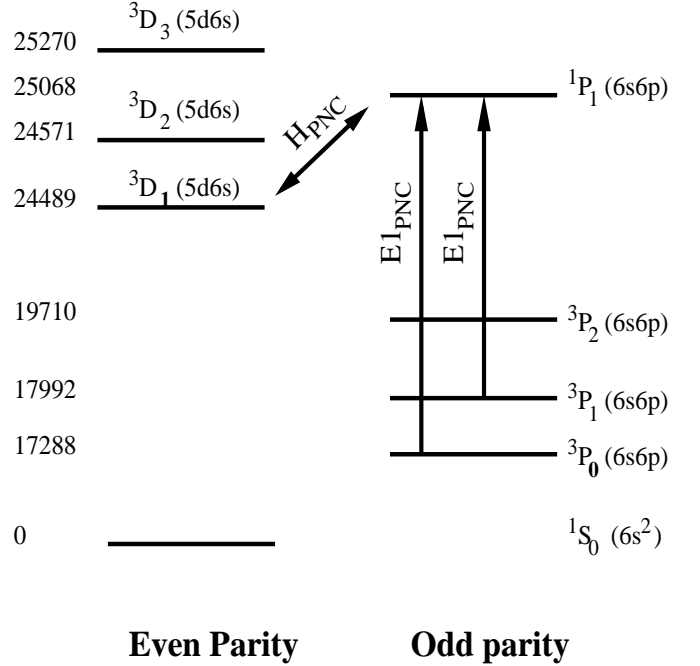}
   \caption{\label{figure1}
            The low-lying levels of atomic Yb. The double arrow represents the 
            $H_{\rm PNC}$ mixing and single arrow represents the transitions
            of interest for $E1_{\rm PNC}$ measurement.} 
\end{figure}

  The low-lying levels of the Yb are given in the Fig \ref{figure1}. The 
$H_{\rm PNC}$ mixing between the $|^1P_1(6s6p)\rangle $ and 
$|^3D_1(5d6s)\rangle$ reduces to mixing between $5d$ and $6p$ at the single 
particle level, which is negligible as $5d$ is almost zero in the nuclear 
region.  However, $6s$ and $6p$ mixing can occur through the  mixing
$\langle ^3D_1(5d6s)|H_{\rm PNC}|5d6p\rangle$, which arises due to the strong 
configuration mixing between $|6s6p\rangle$ and $|5d6p\rangle$. So the most 
important contribution to the $E_{\rm PNC}(^3P_0 - ^1P_1)$ transition 
amplitude is
\begin{eqnarray}
   & & \frac{\langle^1P_1(6s6p)|D|5d6p\rangle\langle 5d6p|H_{\rm PNC}
       |^3P_0(6s6p) \rangle}{E_{|^3P_0(6s6p)\rangle} - E_{|5d5p\rangle}} 
                                                   \nonumber \\
   &+& \frac{\langle^1P_1(6s6p)|H_{\rm PNC}|5d6p\rangle\langle 5d6p|D
       |^3P_0(6s6p) \rangle}{E_{|^1P_1(6s6p)\rangle} - E_{|5d5p\rangle}}.
                                                   \nonumber 
  \label{mjrcnt}
\end{eqnarray}
This indicates that the choice of $|^3P_1(6s6p)\rangle$ as initial state
offers the possibility of non-zero contribution from the leading
configurations to the normal  and conjugate terms. Whereas the largest
contribution to the $E_{\rm PNC} (^1S_0(6s^2) - ^1P_1(6s6p) )$ is
\begin{equation}
   \frac{\langle^1P_1(6s6p)|D|5d6p\rangle\langle 5d6p|H_{\rm PNC}
       |^1S_0(6s^2) \rangle}{E_{|^1S_0(6s6p)\rangle} - E_{|5d5p\rangle}},
\end{equation}
the conjugate term does not contribute as $|5d6p\rangle$ is double excitation
with respect to $|^1S_0(6s^2)\rangle$.

  All the calculations are done using GRASP2\cite{dyall-425-89} package.
The choice of the screening parameters $\alpha$s and the dependence of the
excitation energies on these parameters are discussed in an earlier 
work\cite{das-1635-97,angom-4905-99}. The values of the $\alpha_K$s which can 
give excitation energies very close to the experimental data are 
$\alpha_0 = 0.997$, $\alpha_1 = 0.667$, $\alpha_2 = 0.980$ and rest are set 
to unity. The even and odd parity CSF space consist of the following 
non-relativistic configurations: \\
\noindent Even parity
\begin{eqnarray}
   & & \left ( 5p^64f^{14}\right ) \left ( 6s^2   + 5d6s   + 5d^2   + 6p^2) 
                                   \right ) \nonumber \\ 
   &+& \left ( 5p^64f^{13}\right ) \left ( 6s^26p + 5d^26p + 5d6s6p         
                                   \right ) \nonumber \\ 
   &+& \left ( 5p^54f^{14}\right ) \left ( 6s^26p + 5d^26p + 5d6s6p        
                                   \right ) \nonumber 
\end{eqnarray} 
\noindent Odd parity
\begin{eqnarray}
   & & \left ( 5p^64f^{14} \right ) \left ( 6s6p   + 5d6p 
                                    \right ) \nonumber \\ 
   &+& \left ( 5p^64f^{13} \right ) \left ( 5d^26s + 5d6s^2 + 6s6p^2 + 5d6p^2 
                                    \right ) \nonumber \\ 
   &+& \left ( 5p^54f^{14} \right ) \left ( 5d^26s + 5d6s^2 + 6s6p^2 + 5d6p^2 
                                    \right ) \nonumber 
\end{eqnarray}
The configuration space is chosen to represent the important valence-valence
and core-valence correlation  effects.  The excitation energies calculated
using these choices of $\alpha_K$s and configurations are given in 
Table\ref{table1}.
\begin{table}[h]
   \caption{\label{table1}
            The excitation energies of the low-lying levels in $cm^{-1}$ } 
   \begin{ruledtabular}
   \begin{tabular}{ccc}
      Level     & Expt. data   &  Theory  \\ \hline
     $^3P_1 (6s6p)$ &      17992          &       17651    \\
     $^3P_2 (6s6p)$ &      19710          &       19855    \\
     $^3D_1 (5d6s)$ &      24489          &       24467    \\
     $^1P_1 (6s6p)$ &      25068          &       25024    \\
   \end{tabular}
   \end{ruledtabular}
\end{table} 
For a comparative study we have also calculated 
$E1_{\rm PNC}(^3P_1(6s6p) - ^1P_1(6s6p))$. However $|^3P_1(6s6p)\rangle$ can 
decay to the ground state $|^1S_0(6s^2)\rangle$ by magnetic quadrupole 
transition. Where as $|^3P_0(6s6p)\rangle - |^1S_0(6s^2)\rangle$ is highly 
forbidden. The calculated values of reduced matrix element of 
$E1_{\rm PNC}$ are given in 
Table \ref{table2}.
\begin{table}[h]
   \caption{\label{table2} The $E1_{\rm PNC}$ reduced matrix elements of atomic
            $^{171}{\rm Yb}( I = 1/2)$ in units of 
            $iea_0 (-Q_W/N) \times 10 ^{-11}$.}
   \begin{ruledtabular}
   \begin{tabular}{cc}
     Transition                            &  $||E1_{\rm PNC}||$\\   \hline
     $^1S_0(6s^2) - ^3D_1(5d6s)$  &   77.8         \\
     $^3P_1(6s6p) - ^1P_1(6s6p)$  &  -71.7         \\
     $^3P_0(6s6p) - ^1P_1(6s6p)$  &  -96.0         \\
   \end{tabular}
   \end{ruledtabular}
\end{table}
For comparison $||E1_{\rm PNC}(^1S_0(6s^2) - ^3D_1(5d6s))||$ is also included 
in the table. The $||E1_{\rm PNC}||$ of atoms and ions which has been studied 
and calculated recently are given in Table \ref{table3}.
\begin{table}[h]
   \caption{\label{table3} $E1_{\rm PNC}$ of recently studied and calculated
            atomic/ionic transitions. These are given in units of 
            $iea_0(-Q_W/N)\times 10^{-11}$ }
   \begin{ruledtabular}
   \begin{tabular}{ccclll}
    Atom/ion & Z  &  N  &  \multicolumn{2}{c}{Transition}  &  
           $E1_{\rm PNC}$ \\ \cline{4-5}
                &    &     & Initial state      & final state & \\ \hline \hline
        Cs      & 55 &  78 &  $|6s_{1/2}   \rangle$ & $|7s_{1/2}   \rangle$ &
           -0.8991(36)\footnotemark[1]        \\
        Cs      & 55 &  78 &  $|6s_{1/2}   \rangle$ & $|5d_{3/2}   \rangle$ &
            3.75      \footnotemark[2]        \\
        Ba$^+$  & 55 &  82 &  $|6s_{1/2}   \rangle$ & $|5d_{3/2}   \rangle$ &
            2.17      \footnotemark[2]        \\
        Yb      & 70 & 101 &  $|^1S_0(6s^2)\rangle$ & $|^3D_1(5d6s)\rangle$ &
           79.38      \footnotemark[3]        \\
        Tl      & 81 & 124 &  $|6p_{1/2}   \rangle$ & $|6p_{3/2}   \rangle$ &
           66.7$\pm$ 1.7\footnotemark[4]      \\
        Fr      & 87 & 136 &  $|7s_{1/2}   \rangle$ & $|6d_{3/2}   \rangle$ &
           57.1       \footnotemark[2]        \\
        Ra      & 88 & 139 &  $|^1S_0(7s^2)\rangle$ & $|^3D_1(7s6d)\rangle$ &
          77.0       \footnotemark[5]         \\
        Ra      & 88 & 135 &  $|^1S_0(7s^2)\rangle$ & $|^3D_1(7s6d)\rangle$ &
          76.0       \footnotemark[5]         \\
        Ra$^+$  & 87 & 135 &  $|7s_{1/2}   \rangle$ & $|6d_{3/2}   \rangle$ &
          42.9       \footnotemark[2]         \\  
  \end{tabular}
  \end{ruledtabular}
  \footnotetext[1]{See ref\cite{derevianko-1618-2000}}
  \footnotetext[2]{See ref\cite{dzuba-062101-2001}}
  \footnotetext[3]{See ref\cite{das-1635-97}}
  \footnotetext[4]{See ref\cite{kozlov-052107-2001}}
  \footnotetext[5]{see ref\cite{dzuba-062509-2000}}
\end{table}
The comparison of the $E1_{\rm PNC}$ transition amplitudes indicates that
the $^3P_0(6s6p)-^1P_1(6s6p)$ transition of Yb is the largest.

%%%%%%%%%%%%%%%%%%%%%%%%%%%%%%%%%%%%%%%%%%%%%%%%%%%%%%%%%%%%%%%%%%%%%%%%%%%%%%%%
%%%%                            SECTION:3.2                                 %%%%
%%%%                   Results: MCDF calculations                           %%%%
%%%%%%%%%%%%%%%%%%%%%%%%%%%%%%%%%%%%%%%%%%%%%%%%%%%%%%%%%%%%%%%%%%%%%%%%%%%%%%%%

\subsection{MCDF Calculations}

   The occupied orbitals $(1-6)s$, $(2-5)p$, $(3-4)d$ and $4f$, and the 
valence orbitals $6s$, $5d$ and $6p$ are generated by a  sequence of MCDF 
calculations and non of the orbitals are frozen in each of the calculations. 
The non-relativistic configurations of the final calculation of the sequence 
are: $6s^2$, $6p^2$, $5d6s$, $5d^2$, $6s6p$ and $5d6p$. These orbitals are 
generated spectroscopic, that is the number of nodes satisfy the central field 
condition. This selection of non-relativistic configurations include the 
leading configurations of the $^3P_1(6s6p)$, $^3D_1(5d6s)$ and $^1P_1(6s6p)$ 
levels, and the configurations which mix strongly with these. The levels 
obtained as a result of the MCDF calculation are given in the column MCDF1
of Table\ref{table4}.
\begin{table}[h]
  \caption{\label{table4} The Yb levels in units of ${\rm cm} ^{-1}$ 
           calculated using MCDF method with the leading and next leading 
           configurations}
  \begin{ruledtabular}
  \begin{tabular}{cccc}
     Levels  &\multicolumn{3}{c}{Excitation Energies(E) } \\ \cline{2-4} 
              % \multicolumn{2}{c}{$\Delta E$\footnote{$\Delta E = 
              %  E(Expt) - E(MCDF)$}} \\ \cline{2-4} % \cline{5-6}
             & Expt  & MCDF1 & MCDF2 \\ \hline % & MCDF1  &  MCDF2   \\
     $^3P_0$ & 17288 & 13428 & 13535 \\        % &  22.33 &  21.71        \\
     $^3P_1$ & 17992 & 14119 & 14234 \\        % &  21.53 &  20.89        \\
     $^3P_2$ & 19710 & 15644 & 15808 \\        % &  20.63 &  19.80        \\
     $^3D_1$ & 24489 & 25458 & 24784 \\        % &   3.96 &   1.20        \\
     $^3D_2$ & 24571 & 25507 & 24802 \\        % &   3.81 &   0.94        \\
     $^3D_3$ & 25270 & 25591 & 24844 \\        % &   1.27 &   1.68        \\
     $^1P_1$ & 25068 & 24990 & 25078 \\        % &   0.31 &   0.03        \\
  \end{tabular}
  \end{ruledtabular}
\end{table}
The calculated $E(MCDF)$ differ from the experimental data in the range
$0.3\% - 22.0\%$ and the sequence is incorrect as $^1P_1$ lies below the
$^3D_J$. A correct sequence of the levels can be obtained by saturating the
valence correlation effects by including the virtual orbitals $6d$ and $6f$ 
generated as correlation orbitals. The results of the calculation are give in 
column MCDF2 of Table\ref{table4}. The $^3D_j$ and $^1P_1$ are in good 
agreement with the experimental data and the $^1P_1$ lies above $^3D_1$ and 
$^3D_2$ levels. However, it is also above $^3D_3$, which is in disagreement
with the experimental data. This indicates that the correlation orbitals 
$6d$ and $6f$ lower the $^3D_J$ levels but has little effect on the $^3P_J$
and $^1P_1$ levels.

  The MCDF calculation captures the important valence-valence correlation 
effects. However, it is not suitable to capture the core-valence and 
core-core correlation effects, which require a large number of CSFs. CI 
calculations within the CSF space having excitations from the core and valence 
shells to the virtual can capture these correlation effects. The virtual
orbitals required for the CI calculation is generated in layers, where one
layer is a set of orbitals of $s$, $p$, $d$, $f$ and $g$ symmetries having
same principal quantum number. Higher angular momentum orbitals $h$ and above 
are not included in the calculation. Each layer is generated by an MCDF-EOL 
caculation of the CSFs used in the previous calculation and the CSFs obtained
by single excitation from the configurations $6s^2$, $5d6s$, $6s6p$, $6p^2$, 
$6p6f$ and $5d6p$.

\begin{table}[h]
   \caption{\label{table5}The excitation energies of the low-lying levels and 
            the magnetic hyperfine structure constants.}
   \begin{ruledtabular}
   \begin{tabular} {cclll}
             & \multicolumn{2}{c}{Excitation energies} 
             & \multicolumn{2}{c}{Hyperfine Constant A}\\ \cline{2-3}\cline{4-5}
     Level   &  Expt  &  Theory & Expt         & Theory              \\ \hline
     $^3P_0$ &  17288 &  15693  &              &                          \\ 
             &        &  17282\footnotemark[1] &                          \\ 
             &        &  17359\footnotemark[2] &                          \\ 
     $^3P_1$ &  17992 &  16370  & -1094.0(7)\footnotemark[3] & -1065      \\ 
             &        &  17997\footnotemark[1] &  & -1094\footnotemark[1] \\ 
             &        &  18089\footnotemark[2] &  &                       \\ 
     $^3P_2$ &  19710 &  17989  & -738      \footnotemark[4] &  -747      \\ 
             &        &  19750\footnotemark[1] &  &  -745\footnotemark[1] \\ 
             &        &  19836\footnotemark[2] &  &                       \\ 
     $^3D_1$ &  24489 &  23468  &  563      \footnotemark[3] &   601      \\ 
             &        &  24441\footnotemark[1] &  &  596 \footnotemark[1] \\ 
             &        &  24936\footnotemark[2] &  &                       \\ 
     $^3D_2$ &  24571 &  23567  &  -362(2)  \footnotemark[3] &  -307      \\ 
             &        &  24697\footnotemark[1] &  &  -351\footnotemark[1] \\ 
             &        &  25180\footnotemark[2] &  &                       \\ 
     $^3D_3$ &  25270 &  23744  &  -430(1)  \footnotemark[3] &  -454      \\ 
             &        &  25247\footnotemark[1] &  &  -420\footnotemark[1] \\ 
             &        &  25676\footnotemark[2] &  &                       \\ 
     $^1P_1$ &  25068 &  24430  &   59      \footnotemark[4] &  152       \\ 
             &        &  25074\footnotemark[1] &  &   191\footnotemark[1] \\ 
             &        &  27271\footnotemark[2] &  &                       \\ 
   \end{tabular}
   \footnotetext[1]{See ref\cite{rakhlina-9810011-98}}
   \footnotetext[2]{See ref\cite{eliav-2765-95}}
   \footnotetext[3]{See ref\cite{topper-233-97}}
   \footnotetext[4]{See ref\cite{jin-2896-91}}
   \end{ruledtabular}
\end{table}

  The results of a CI calculation within the CSF space spanned by all possible 
excitations from the valence shells and single excitations from the core-shells
$5s$, $5p$ and $4f$ with respect to the $6s^2$, $5d6s$, $6s6p$, $6p^2$, $6p6f$ 
and $5d6p$ are given in Table\ref{table5}. The results of earlier calculations
are also given in the table. The sequence of the excitation energies are 
correct but the diviation from the experimental values is still large. However, 
the level $^1P_1$ is better than the coupled-cluster results and the hyperfine 
constant of the same level is in better agreement with the experimental data. 
The important reduced matrix elements of the electric dipole are given in the 
Table:\ref{table6}.

\begin{table}
   \caption{\label{table6} }
   \begin{ruledtabular}
   \begin{tabular}{cccc}
                  &  $^3P_0(6s6p)$    & $^3P_1(6s6p)$           & 
       $^1P_1(6s6p)$           \\ \hline
    $^1S_0(6s^2)$ &                   & 0.51                    & 
       4.98                    \\
            &                         & 0.54(8)\footnotemark[1] & 
       4.40(80)\footnotemark[1] \\
            &                         & 0.44   \footnotemark[2] & 
       4.44(80)\footnotemark[2] \\
            &                         & 0.549(4)\footnotemark[5]& 
       4.89(80)\footnotemark[3] \\
            &                         &0.553(13)\footnotemark[6]& 
       4.13(10)\footnotemark[6] \\
            &                         &                         & 
       4.02    \footnotemark[7] \\
            &                         &                         & 
       4.26    \footnotemark[8] \\
    $^3D_1(5d6s)$ &3.03               & 2.64                    & 
       0.28      \\
            &2.61(10)\footnotemark[1] & 2.26(10)\footnotemark[1]& 
       0.27(10)\footnotemark[1] \\
            &                         & 2.2(1)  \footnotemark[4]& 
       0.24    \footnotemark[2] \\
   \end{tabular}
   \footnotetext[1]{See ref\cite{porsev-2781-99}}
   \footnotetext[2]{See ref\cite{migdalek-99-91}}
   \footnotetext[3]{See ref\cite{kunisz-285-82}}
   \footnotetext[4]{See ref\cite{bowers-3103-96}}
   \footnotetext[5]{See ref\cite{bowers-3513-99}}
   \footnotetext[6]{See ref\cite{baumann-283-66}}
   \footnotetext[7]{See ref\cite{penkin-78}}
   \footnotetext[8]{See ref\cite{andersen-257-75}}
   \end{ruledtabular}
\end{table}
The $E1_{\rm PNC}$ calculated using the theoretical results is 
$-86.01 \times 10^{-11} i(-Q_W/N) ea_0$. The important contributions are from 
the intermediate states $^3D_1(5d6s)$ and $^1S_0(6s^2)$, each of these yield
$-90.62 \times 10^{-11}(-Q_W/N)ea_0$ and $24.74 \times 10^{-11}i(Q_W/N)ea_0$ 
respectively. The cancellation due to the $^1S_0(6s^2)$ is large but is nearly 
compensated by the contributions from the intermediate states $^3S_1(6s7s)$ 
and $^3P_1(6p^2)$, which totals to $-21.78 \times 10^{-11}i(Q_W/N)ea_0$. 
However, as mentioned earlier, the energy spacing of the important levels are 
not correct. In addition $<^3\!\!P_0(6s6p)||D||^3D_1(5d6s)\!\!>$, 
$<^3\!\!P_0(6s6p)||D||^3S_1(6s7s)\!\!>$ and $|<1S0||D||1P1>|$ also show 
variations from the experimental data and previous theoretical 
calculation\cite{porsev-2781-99}. Using the experimental excitation energies, 
the average of the experimental data of $|<1S0||D||1P1>|$, and the other 
$|D_{fi}|$s calculated by Porsev~et~al\cite{porsev-2781-99} gives 
$-12.31 \times 10^{-10}i(Q_W/N)ea_0$. 

   The contribution from the intermediate state $^3D_1(5d6s)$ 
is $-13.09 \times 10^{-10}i(Q_W/N)ea_0$  which is calculated using the 
reduced matrix element of 
$<^3\!\!D_1(5d6s)||H_{\rm PNC}^{\rm NSI}||^3P_0(6s6p)\!\!>$ 
from the present work, the $<^1\!\!P_1(6s6p)||D||^3P_0(6s6p)\!\!>$ from 
Porsev~et~al\cite{porsev-2781-99} and the enrgy denominator from the
experimental  data. The estimate of the
 $<^3\!\!D_1(5d6s)||H_{\rm PNC}^{\rm NSI}||^3P_0(6s6p)\!\!>$ using the 
wave functions we have calculated is expected to be better as the 
magnetic dipole hyperfine structure constant of $^1P_1(6s6p)$ is in better
agreement with the experimental data.

%%%%%%%%%%%%%%%%%%%%%%%%%%%%%%%%%%%%%%%%%%%%%%%%%%%%%%%%%%%%%%%%%%%%%%%%%%%%%%%%
%%%%                            SECTION:4                                   %%%%
%%%%                           Conclusions                                  %%%%
%%%%%%%%%%%%%%%%%%%%%%%%%%%%%%%%%%%%%%%%%%%%%%%%%%%%%%%%%%%%%%%%%%%%%%%%%%%%%%%%

\section{Conclusions}

  Our MCDF calculations shows that the 
$E1_{\rm PNC}(^3P_0(6s6p) - ^1P_1(6s6p))$ PNC transition amplitude of 
$^{171}{\rm Yb}$ is more than two order of magnitude larger than that of the 
$E1_{\rm PNC} ( 6s - 7s)$. Our calculation indicates the experimental 
proposal of Kimball\cite{kimball-052113-2001} shows that 
$^3P_0(6s6p)-^1P_1(6s6p)$ is indeed a promising candidate for the 
$E1_{\rm PNC}$ measurements.

%%%%%%%%%%%%%%%%%%%%%%%%%%%%%%%%%%%%%%%%%%%%%%%%%%%%%%%%%%%%%%%%%%%%%%%%%%%%%%%%
%%%%                            SECTION:4                                   %%%%
%%%%                        Acknowledgements                                %%%%
%%%%%%%%%%%%%%%%%%%%%%%%%%%%%%%%%%%%%%%%%%%%%%%%%%%%%%%%%%%%%%%%%%%%%%%%%%%%%%%%

\section{Acknowledgements}

   These calculations were done using the Enterprise450 at the Indian 
Institute of Astrophysics, Bangalore and the IBM workstations at the  Physical
Research Laboratory, Ahmedabad.

%%%%%%%%%%%%%%%%%%%%%%%%%%%%%%%%%%%%%%%%%%%%%%%%%%%%%%%%%%%%%%%%%%%%%%%%%%%%%%%%
%%%%                            SECTION:4                                   %%%%
%%%%                           Bibliography                                 %%%%
%%%%%%%%%%%%%%%%%%%%%%%%%%%%%%%%%%%%%%%%%%%%%%%%%%%%%%%%%%%%%%%%%%%%%%%%%%%%%%%%

\end{document}